\documentclass[prb,
superscriptaddress,showpacs,amsmath,amssymb]{revtex4}
\usepackage{amsfonts}
\usepackage{bm}
\usepackage{verbatim}

\usepackage{graphicx}

\newcommand{\id}{\mathbf{1}}

\begin{document}

\title{Type-III and IV interacting Weyl points}

\author{J.~Nissinen}
\affiliation{Low Temperature Laboratory, Aalto University,  P.O. Box 15100, FI-00076 Aalto, Finland}

\author{G.E.~Volovik}
\affiliation{Low Temperature Laboratory, Aalto University,  P.O. Box 15100, FI-00076 Aalto, Finland}
\affiliation{Landau Institute for Theoretical Physics, acad. Semyonov av., 1a, 142432,
Chernogolovka, Russia}

\date{\today}

\begin{abstract}
3+1-dimensional Weyl fermions in interacting systems are described by effective quasi-relativistic Green's functions parametrized by a 16 element matrix $e^\mu_\alpha$ in an expansion around the Weyl point. The matrix $e^{\mu}_{\alpha}$ can be naturally identified as an effective tetrad field for the fermions. The correspondence between the tetrad field and an effective quasi-relativistic metric $g_{\mu\nu}$ governing the Weyl fermions allows for the possibility to simulate different classes of metric fields emerging in general relativity in interacting Weyl semimetals. According to this correspondence, there can be four types of Weyl fermions, depending on the signs of the components $g^{00}$ and $g_{00}$ of the effective metric. In addition to the conventional type-I fermions with a tilted Weyl cone and type-II fermions with an overtilted Weyl cone for $g^{00}>0$ and respectively $g_{00}>0$ or $g_{00}<0$, we find additional ``type-III" and ``type-IV" Weyl fermions with instabilities (complex frequencies) for $g^{00}<0$ and $g_{00}>0$ or $g_{00}<0$, respectively. While the  type-I and type-II Weyl points allow us to simulate the black hole event horizon at an interface where $g^{00}$ changes sign, the type-III Weyl point leads to effective spacetimes with closed timelike curves.
\end{abstract}
\pacs{
}

\maketitle

\section{Introduction}

Weyl fermions \cite{Weyl1929} are massless fermions whose masslessness (gaplessness) is topologically protected \cite{NielsenNinomiya1981,FrogNielBook,GrinevichVolovik1988,Horava2005}.
They appear as the building blocks of the Standard Model (SM) of fundamental interactions and as quasi-relativistic quasiparticles in topological semimetals \cite{Herring1937,Abrikosov1971,Abrikosov1972, WanEtAl2011, Burkov2011a, Weng2015, Huang2015,Lv2015,Xu2015,Lu2015, SunEtAl2015, ChangEtAl2016, XuEtAl2016b, XuEtAl2016, KoepernikEtAl2016, WangEtAl2016, TamaiEtAl2016, JiangEtAl2016, HuangZhouDuan2016, LeEtAl2016, ChangEtAl2016b} and chiral superfluid $^3$He-A \cite{Bevan1997,Volovik2003}.
In terms of the energy spectrum of the single-particle Hamiltonian, the Weyl point is a topologically protected point where two non-degenerate bands touch each other with linear dispersion \cite{NeumannWigner1929,Novikov1981}. In the vicinity of the exceptional band touching, the spectrum is described by a $2\times 2$ effective Weyl Hamiltonian, which incorporates only the two touching bands. The most general $2\times 2$ Hamiltonian $H(\mathbf{p}) = e^{i}_{\alpha}\sigma^{\alpha}p_i$ contains 12 parameters $e^{i}_{\alpha}$ which correspond to the elements of a tetrad field in general relativity (GR).  Depending on the relations between these parameters, the Weyl fermions associated with the Weyl point can be divided into two distinct classes \cite{VolovikZubkov2014, YongXu2015} called type-I and type-II \cite{Soluyanov2015}. In the type-II case, the Weyl point is the topologically protected conical point in momentum space, which connects the electron and hole pockets in the semimetal. In GR such a fermionic spectrum can emerge behind a black hole horizon \cite{HuhtalaVolovik2002,Volovik2016}.

Here we suggest that for the interacting fermions the situation can be more complicated. In the vicinity of the Weyl point in 4-momentum space $p_{\mu} = (\omega, \mathbf{p})$, the expansion of the Green's function contains all 16 components $e^{\mu}_{\alpha}$ of the tetrad field. This leads to two additional distinct classes of Weyl points, type-III and type-IV, depending on the four signs of the elements $g^{00}$ and $g_{00}$ of the effective metric. This allows us to simulate other specific solutions of Einstein equations at interfaces where the components of the metric change sign, which includes  spacetimes with closed timelike curves.

 \section{General Weyl points for interacting fermions}
 \label{tetrads}
 
The general $2\times 2$ matrix Hamiltonian describing Weyl fermions in the vicinity of the Weyl point at ${\bf p}=0$ has the form
\begin{equation}
H({\bf p})= e_0^i \sigma^0 p_i  +  e_a^i \sigma^a p_i \,,
\label{HamiltonianWeyl}
\end{equation}
where $p_i$ with $i=x,y,z$ are the components of three-momentum $\mathbf{p}$, $\sigma^0=\id$ and $\sigma^a$ for $a=1,2,3$ are the Pauli matrices. Recently, the Weyl points with overtilted cones due to non-zero vector coefficient $e_0^i \sim \mathbf{v}$ were dubbed type II Weyl fermions \cite{Soluyanov2015}.
The coefficients in the linear expansion -- the vector $e_0^i$ and the matrix $e_a^i$ -- are equivalent to the components of the tetrad field $e^\mu_\alpha$ in relativistic theories, where $\mu,\alpha=0,1,2,3$ in  3+1 dimensions. More specifically, in the first order formalism of general relativity, a given metric is expressed in terms of the tetrad field as
\begin{equation}
 g^{\mu\nu}=\eta^{\alpha\beta}e^\mu_\alpha e^\nu_\beta 
 \,,
\label{EffectiveMetricTetrad}
\end{equation}
where the (dimensionless) metric $\eta^{\alpha\beta}={\rm diag}(1,-1,-1,-1)$ has Lorentzian signature. In general, the fermions couple to the tetrad and the metric $g_{\mu\nu}$ even in flat space, since the gamma matrices in the background metric \eqref{EffectiveMetricTetrad} are given as
\begin{align}
\gamma^{\mu} = e^{\mu}_\alpha \gamma^{\alpha}, \label{eq:4x4gamma}
\end{align}
and satisfy the Clifford anticommutation relations $\{\gamma^{\mu}, \gamma^{\nu}\} = 2g^{\mu\nu}$ with $\gamma^0 = \tau_1\otimes \id, \gamma^{a} = i\tau_2 \otimes \sigma^{a}$ for $\alpha=0,a$, and $\tau_i$ are Pauli matrices in the chiral basis, labeling two Weyl points of opposite chirality. In contrast to the full $4\times4$ Dirac representation in Eq. \eqref{eq:4x4gamma}, here we focus on a single two-component Weyl fermion with Hamiltonian Eq. \eqref{HamiltonianWeyl} around a Weyl point. 

Neither the Weyl Hamiltonian Eq. \eqref{HamiltonianWeyl} nor the Dirac Hamiltonian contain all the components of the tetrad field: the elements $e^0_a$ and $e^0_0$ are missing. This is because in the Hamiltonian description the zero component $p_0$ of the four vector $p_\mu = (\omega, \mathbf{p})$ is absent. Nevertheless, the $p_0$ component appears in the Green's function formalism, since the Green's function $G(\omega,{\bf p})$ depends on the frequency $\omega\equiv p_0$. 
For noninteracting fermions, the Green's function
\begin{equation}
G^{-1}(\omega,{\bf p})=    H  - \omega = e^0_0 \sigma^0 p_0+ e_0^i \sigma^0 p_i  +  e_a^i \sigma^a p_i \,.
\label{GreensWeyl}
\end{equation}
contains the term $e^0_0$ with fixed $e^0_0=-1$, but the term with $e^0_a$  is still missing. This term may appear in a interacting system, e.g. from the fermion self-energy,  where the general form of the $2\times 2$ Green's function is:
\begin{equation}
G^{-1}(p_\mu)=    e^0_0 \sigma^0 p_0+ e_0^i \sigma^0 p_i  +  e_a^i \sigma^a p_i +  e_a^0 \sigma^a p_0\,.
\label{GreensWeyGeneral}
\end{equation}
The effective metric Eq. \eqref{EffectiveMetricTetrad} for the Weyl fermions is obtained by considering the poles of the Green's function, which are given by the equation
\begin{equation}
 (e^0_0p_0+ e_0^i p_i)^2 -({\bf e}^i  p_i + {\bf e}^0  p_0)^2=0
   \,\, , \,\, \, 
   {\bf e}^i \equiv e_a^i 
     \,\, , \,\, \, \,
       {\bf e}^0 \equiv e_a^0 \,.
\label{GreensPoles}
\end{equation}
The dispersion relation Eq.(\ref{GreensPoles}) can be written in terms of the metric $g^{\mu\nu}$ as
\begin{equation}
 g^{\mu\nu}p_\mu p_\nu=0
 \,,
\label{EffectiveMetricTetrad1}
\end{equation}
and leads to Eq.(\ref{EffectiveMetricTetrad}) for both Weyl and massless Dirac fermions, since the metric cannot resolve between the spinor representations. Equivalently, the symmetric matrix $g^{\mu\nu}$ defines a quadratic surface in 4-momentum space which is conical for the quasi-relativistic Lorentz signature $\eta^{\alpha\beta}$. We note that in our conventions we take $\eta^{\alpha\beta}$ to be dimensionless, so $e^0_{\alpha}$ is dimensionless, while $e^{i}_{\alpha}$ carries the dimensions of velocity.

\subsection{Effective interacting Weyl geometry}
While the coefficients $e_a^0$ do not enter the single-particle Weyl Hamiltonian Eq. \eqref{HamiltonianWeyl}, the emergence of the tetrad elements $e^0_a$ in an interacting system may drastically change the behavior of the fermions. First, the matrix element
\begin{equation}
 g^{00}= \left(e^0_0 \right)^2-  \left({\bf e}^0\right)^2
 \,,
\label{g00}
\end{equation}
in the fermion dispersion relation now may cross zero and become negative. Second, the matrix element $g^{0i}$ contains the two tetrad vectors
$e^0_a$ and $e_0^i$:
\begin{equation}
 g^{0i}= e^0_0 e_0^i  -  {\bf e}^0 \cdot  {\bf e}^i 
 \,.
 \label{g0i}
\end{equation}
Let  us consider the effect of these two vectors in the simplest spatially isotropic case, i.e. when the matrix $e^i_a$ is isotropic and characterized by the ``speed of light" or Fermi-velocity $c$ of an isotropic Weyl cone. In this case we can parametrize
\begin{equation}
 e^0_0= -1
   \,\, , \,\, \, 
  e^i_a= c \delta^i_a
     \,\, , \,\, \, 
  e^i_0= v^i     \,\, , \,\, \, \,
       e^0_a=\frac{w_a}{c}  \,.
\label{SimplestTetrad}
\end{equation}
Both tetrad vectors $e_0^i, e_a^0$ can be characterized in terms of the two velocity fields ${\bf v}$ and ${\bf w}$, respectively, from which the effective metric $g^{\mu\nu} = g^{\mu\nu}(\mathbf{v},\mathbf{w})$ for the interacting Weyl point follows as
\begin{align}
 g^{00}(\mathbf{v},\mathbf{w}) = 1-\frac{w^2}{c^2}
   \,\, , \,\, \, 
   g^{0i}(\mathbf{v},\mathbf{w})= -(w^i  +v^i) 
     \,\, , \,\, \, 
  g^{ij}(\mathbf{v},\mathbf{w})= - c^2\delta^{ij} +v^i v^j  
 \,,
\label{SimplestMetric}
\end{align}
with the determinant
\begin{align}
 g(\mathbf{v},\mathbf{w}) &\equiv \det g_{\mu\nu} =-\frac{1}{e^2} = -\frac{1}{c^2 (c^2+{\bf v}\cdot{\bf w})^2}  \,\,\, , \,\, \, \\
 e&\equiv\det e_\alpha^\mu=-c(c^2+{\bf v}\cdot{\bf w})  \,.
\label{SimplestDeterminant}
\end{align}
These enter the equation for the poles in the Green's function in the following way:
\begin{equation}
 (\omega-{\bf v}\cdot {\bf p})^2 -c^2 \left({\bf p} + \frac{{\bf w}}{c^2}\omega\right)^2=0\,,
\label{SimplestSpectrum}
\end{equation}
or 
\begin{equation}
  \omega^2\left(1 - \frac{w^2}{c^2}\right)  - 2\omega({\bf w}+{\bf v})\cdot {\bf p} -c^2(p^2 - (\frac{{\bf v}}{c^2} \cdot {\bf p})^2)=0\,.
\label{SimplestSpectrum2}
\end{equation}
It follows that the dispersions of the two bands touching at the Weyl point are given as
\begin{align}
\omega_{\pm}(\mathbf{p}) =  \frac{-g^{0i}p_i \pm \sqrt{g^{00}(-g^{ij}p_i p_j)+(g^{0i}p_i)^2}}{g^{00}} \hspace*{2cm} \nonumber\\
= \frac{\mathbf{p}\cdot (\mathbf{v}+\mathbf{w})\pm \sqrt{(1-\frac{w^2}{c^2})(c^2 p^2 -(\mathbf{p}\cdot \mathbf{v})^2)+(\mathbf{p}\cdot(\mathbf{v}+\mathbf{w}))^2}}{1-\frac{w^2}{c^2}}.
\label{eq:WeylSpectrum}
\end{align}
From these equations, we see that the velocity field $\mathbf{w}$ is the ``time-like" equivalent of the Type I-II perturbation or velocity field $\mathbf{v}$. They characterize the poles of Green's function Eq. \eqref{SimplestSpectrum2} almost symmetrically between $\omega$ and $\mathbf{p}$. 

The line-interval or covariant metric $g_{\mu\nu}(\mathbf{v},\mathbf{w})$ corresponding to the inverse of $g^{\mu\nu}(\mathbf{v},\mathbf{w})$ follows directly from Eq. \eqref{SimplestMetric} but its form is not very illuminating for general $\mathbf{v}$ and $\mathbf{w}$. However, the time-like components $g^{00}$ and $g_{00}$ are given as 
\begin{align}
g_{00}(\mathbf{v},\mathbf{w}) = \frac{c^4 (c^2-v^2)}{g}, \quad g^{00}(\mathbf{v},\mathbf{w}) = 1-\frac{w^2}{c^2}
\label{eq:g00s}
\end{align}
and a natural proxy for singular behavior in the Weyl spectrum is when these components of the effective metric change sign. Similar behavior of the metric, i.e. interfaces where $g^{00}, g_{00}$ change sign, also arise for the solutions of GR with \emph{physical} singularities, although in that case the metric, nor any of its components, is not directly physical due to the general coordinate invariance of GR. 

In case of the Weyl metric, for $v^2>c^2$, the components $g_{00}$ and $g^{vv}$ of the metric change sign, whereas $g^{00}$ changes sign for $w^2>c^2$. In the latter case for $\mathbf{w}=0$, the effective velocity of the lower branch in the direction $\mathbf{v}$ changes sign, forming the type-II Weyl point. On the other hand, for $w^2>c^2$ and $\mathbf{v}=0$, both $g^{00}$ and $\omega_{\pm}$ change sign and the frequency becomes complex in general. We will call spectra with $w^2 > c^2$ and $v^2 < c^2$, i.e. $g^{00}<0$ and $g_{00}>0$, type-III Weyl. In addition to this, a distinct type-IV spectrum corresponding to the region with both $g^{00}<0$ and $g_{00}<0$ also emerges.

Lastly, in the limiting case when either of the velocity fields $\mathbf{v}$ and $\mathbf{w}$ vanishes the metric and its inverse acquires a simple form. It turns out that the remaining non-zero perturbation enters the effective Weyl metric $g^{\mu\nu}(\mathbf{v},\mathbf{w})$ in Eq. \eqref{SimplestMetric} similarly as the so-called shift vector $N^i$ in the 3+1-dimensional Hamiltonian or ADM decomposition of general relativity \cite{ADM}, where
\begin{align}
ds^2 = g^{\rm ADM}_{\mu\nu} d x^{\mu} d x^{\nu}= N^2 dt^2 - \tilde{g}_{ij}(dx^i + N^i dt)(dx^j+N^j dt), \label{eq:ADMform}\\
g^{\rm ADM}_{\mu \nu} = \left(\begin{matrix} N^2-N^iN_i & -N_j \\ -N_i & -\tilde{g}_{ij}  \end{matrix}\right), 
g_{\rm ADM}^{\mu\nu} = \left(\begin{matrix} \frac{1}{N^2} & -\frac{N^j}{N^2} \\ -\frac{N_i}{N^2} & \frac{N^i N^j}{N^2} -\tilde{g}^{ij}  \end{matrix}\right). \nonumber
\end{align}
In condensed matter applications, we can fix $N=1$ by a time reparametrization, leaving the spatial metric $\tilde{g}_{ij}$ and $N^i$ as parameters. For $\mathbf{w}=0$, we can directly identify $N^i = v^i$ and $\tilde{g}^{ij} = c^2 \delta^{ij}$ (as expected) from $g^{\mu\nu}(\mathbf{w}=0)$. On the other hand, from $g^{\mu\nu}(\mathbf{v} = 0)$, we identify the matrix form of the ADM \emph{covariant metric} $g^{\rm ADM}_{\mu\nu}$ with $N_i = w^i$ and $\tilde{g}_{ij} = c^{2}\delta_{ij}$ and vice versa. For both $\mathbf{v}, \mathbf{w} \neq 0$ the inverse $g_{\mu\nu}(\mathbf{v},\mathbf{w})$ is more complicated and in-between the two extremes in  \eqref{eq:ADMform} corresponding to $\mathbf{v}=0$ or $\mathbf{w}=0$ .

\section{Singularities in the interacting Weyl spectrum}

\subsection{From type-I to type-II}
 \label{typeII}

The singularity takes place when the velocity $w({\bf r})$ reaches the "speed of light" 
$c({\bf r})$ of the non-interacting system and crosses it.  For $w^2<c^2$, the system reproduces the conventional behavior of Weyl fermions: they behave as type-I Weyl fermions for $v^2<c^2$ and as type II Weyl fermions for $v^2>c^2$. 
The effective interval $d s^2 = g_{\mu\nu} dx^{\mu} dx^{\nu}$ has
\begin{align}
g_{00} = \frac{c^4 (c^2-v^2)}{g}, \quad g^{00} = 1-\frac{w^2}{c^2} \,,
\label{g00}
\end{align}
implying that $g_{00}$ of the effective geometry crosses zero at $v^2 = c^2$. The interface $v^2=c^2$ between semimetals with type-I and type-II Weyl fermions serves as the analog of either the ergosurface or of the horizon of an analog blackhole \cite{Volovik2016},  and now this depends on the orientations of the two velocity vectors $\mathbf{v}, \mathbf{w}$ with respect to the interface $v^2=c^2$.

 \subsection{From type-I to type-III}
 \label{typeIII}

For  $w^2>c^2$ one has $g^{00}<0$ and the frequency of the Weyl fermion becomes complex in general. This is somewhat similar to the surface of the black hole singularity in Fig. 32.10 of Ref. \cite{Volovik2003},
where a bosonic ``ripplon" mode acquires complex frequency, which leads to the vacuum instability behind the singular surface of an effective metric. In the ripplon case, the determinant of the metric  $g$ changes sign. The general relativity example where the determinant of the metric changes sign can be found in Ref. \cite{Frolov2004}.
However, in contrast to the ripplon case, in the fermionic system the determinant of the metric  $g$ in Eq.(\ref{SimplestMetric}) does not change sign, only the metric component $g^{00}$.

The simplest example is when ${\bf v}=0$, where the effective Weyl metric \eqref{SimplestMetric} gives:
\begin{equation}
  ds^2=\left( dt - \frac{1}{c^2}{\bf w}\cdot d{\bf r} \right)^2 -\frac{1}{c^2} d{\bf r}^2\,.
\label{interval_w=0}
\end{equation}
and the dispersions:
\begin{align}
  \omega_\pm(\mathbf{p}) &= \frac{1}{1 - \frac{w^2}{c^2}} \left(  
  {\bf w}\cdot {\bf p} 
  \pm \sqrt{ p^2c^2 - ({\bf w}\times {\bf p})^2  }
  \right) \nonumber\\
  &=\frac{1}{1-\frac{w^2}{c^2}}\left(wp_{\parallel} \pm \sqrt{c^2p_{\parallel}^2+(c^2-w^2)p^2_{\perp}}\right),
  \label{SimplestSpectrum4}
\end{align}
where $p_{\parallel}$ and $p_{\perp}$ are defined with respect to $\mathbf{w}$. For $w<c$ one has the type-I Weyl point with tilted cone. For $w>c$ the energy has both positive and negative imaginary parts for the directions of momentum, at which $|{\bf w}\times \hat{\bf p}|> c$. This manifests the instability of the system or localization in the directions transverse to ${\bf w}$. The momentum space surface separating the propagating fermionic states from the fermionc states with complex spectrum forms the cone $p_\parallel= \pm p_\perp\sqrt{\frac{w^2}{c^2}-1}$. We call such conical point a Weyl point of type-III. Analogously, Weyl fermions with both $w>c$ and $v>c$ can be called type-IV fermions.

\subsection{Lifshitz transition with a change of sign of tetrad determinant}
\label{TetradDeterminantSign}

There is also the Lifshitz transition, which is not manifested in the metric $g^{\mu\nu}$, but is seen in the tetrad field. This happens when the tetrad determinant in Eq.(\ref{SimplestDeterminant}) crosses zero at ${\bf v}\cdot{\bf w}=-c^2$. To see the physical meaning of this transition, let us consider the tetrad field with anisotropic speeds of light:
\begin{equation}
 e^0_0= -1
   \,\, , \,\, \, 
  e^i_a= c_x \hat x^i \hat x_a + c (\hat y^i \hat y_a + \hat z^i \hat z_a)
    \,,
\label{TetradSign}
\end{equation}
and choose the direction of either of the vectors $e^i_0$ and $e^0_a$ along $\hat{\bf x}$.
Then, the tetrad determinant is
\begin{equation}
e=\det e_\alpha^\mu=-c^2(c_x+e^i_0 e^0_i)  \,.
\label{TetradDeterminantSign}
\end{equation}
For $e^i_0 =e^0_i=0$ and  $c_x>0$, $c>0$, the determinant $e<0$ and the Hamiltonian describes the right-handed fermions. At $e^i_0e^0_i =-c_x$ a Lifshitz transition takes place, and in the region where $e^i_0e^0_i +c_x <0$  the determinant becomes positive,  $e>0$. In this region one can continuously change the sign of $c_x$ without the change of sign of $e$. Then pushing  $e^i_0e^0_i$ to zero again, one obtains the diagonal tetrad field with  $c_x<0$, $c>0$ and $e>0$, which corresponds to the left-handed fermions. This means that the Lifshitz transition at ${\bf v}\cdot{\bf w}=-c^2$ in Eq. \eqref{SimplestDeterminant} allows for the transformation of the right-handed fermion to the left-handed one. 
In case of Dirac fermions, this transition would correspond to the interchange of the right- and left-handed components of the Dirac fermion. Without the tetrad components $e^i_0$ or $e^{0}_{a}$, the transition changing the chirality would always encompass a nodal line in the spectrum. 

The continuous change of sign of the determinant of the tetrad field may happen in quantum gravity, and it is argued that because of that the quantum gravity cannot be constructed using the diffeomorphism invariant theories \cite{Diakonov2011,Diakonov2012}.

\subsection{Closed timelike curves}
 \label{CTC}

Using  the Weyl point of type III, one may simulate the closed timelike curves in general relativity. That is, in the instability region, one may have closed timelike curves $f(\tau)$, satisfying $\dot{f}_{\mu}\dot{f}_{\nu}g^{\mu\nu} >0$.  

Let us consider the simplest example with
${\bf v}=0$, the Eq.(\ref{interval_w=0}). If one takes an axially symmetric velocity field ${\bf w}= w(r)\hat{{\mbox{\boldmath$\phi$}}}$, one obtains the following effective line-interval for Weyl fermions:
\begin{align}
  ds^2=dt^2  -\frac{1}{c^2} (dz^2+dr^2) -\frac{1}{c^2} \left(1 - \frac{w^2(r)}{c^2}\right)r^2 d\phi^2 - 2\frac{w(r)}{c^2}r d\phi dt\,.
\label{interval_circular}
\end{align}
 At $w^2(r)>c^2$ one has $g_{\phi\phi}>0$, and the closed timelike curves appear with
$t={\rm const}$, $z={\rm const}$ and $r={\rm const}$:
\begin{equation}
  ds^2= g_{\phi\phi} d\phi^2=\frac{r^2 }{c^4}\left(w^2(r)-c^2 \right) d\phi^2 > 0
\,\,,\,\, w^2(r)>c^2\,.
\label{interval_CTC}
\end{equation}
So, while the interface between 
type-I and type-II Weyl points simulates the black hole horizon or ergoregion\cite{Volovik2016}, the interfaces between 
type-I and type-III Weyl points or  between 
type-II and type-III Weyl points simulate closed timelike curves.

In general relativity the closed timelike curves appear as the solutions of Einstein equations, see e.g. Refs.\cite{Bonnor2005,Kajari2004}. In our case they appear only in the unstable state of semimetal, where 
the element $g^{00}$ in effective metric has wrong sign, and the fermionic vacuum becomes unstable.

\section{Conclusion}
 \label{conclusion}

Interacting Weyl fermions are described by all 16 matrix elements $e^\mu_\alpha$ of the tetrad field entering the Green's function. This allows us to simulate different classes of the metric field in general relativity. According to this correspondence, there can be four types of Weyl fermions, depending on the signs of the effective metric elements 
$g^{00}$ and $g_{00}$ in \eqref{eq:g00s}.
In addition to the conventional type-I fermions with tilted Weyl cone for
 $g^{00}>0$ and $g_{00}>0$ and the type-II fermions with overtilted Weyl cone for $g^{00}<0$ and $g_{00}>0$, there are
 type-III and type IV fermions with complex frequencies for  $g^{00}<0$, $g_{00}>0$ and  $g^{00}<0$, $g_{00}<0$ correspondingly. Moreover, the sign of the determinant of the tetrad field can change at a yet another Lifshitz transition, where the right- and left-handed fermions are interchanged.

While the type-II Weyl point is a topologically protected conical point in momentum space, which in semimetals connects the electron and hole pockets, the type-III Weyl point marks a topologically protected cone which separates the propagating fermions from the fermions with complex spectrum. The transitions between type-I, type-II, type-III  and type-IV Weyl points are peculiar forms of Lifshitz transitions.
The interfaces between the states with different types of Weyl points serve as analogs of the special surfaces in general relativity, such as event horizon, ergosurface. When the  type-III  and type-IV Weyl points are also involved, the interfaces represent physical singularities which cannot be removed by 
coordinate transformations. Behind such singular surfaces, closed timelike curves are possible to exist.

In some cases the topology of interacting fermionic systems can be described using 
the effective \cite{Volovik2009,Volovik2010,VayrynenVolovik2011}
(or topological \cite{Wang2013,Abanin2014}) Hamiltonian, which is represented by the Green's function at zero frequency: 
\begin{equation}
H_{\rm eff}({\bf p}) = G^{-1}(\omega=0,{\bf p}) \,.
\label{EffectiveH}
\end{equation}
This effective Hamiltonian is applicable for description of type-I and type-II Weyl points with $\mathbf{v}\neq0, \mathbf{w}=0$ and can resolve between them. However, it cannot resolve the transition to the  type-III or to the type-IV Weyl point with $\mathbf{w}\neq 0$. This demonstrates the breakdown of the  approach with the topological Hamiltonian in Eq. \eqref{EffectiveH} for the interacting case with non-zero $e^0_{a} \sim \mathbf{w}$ in the Green's function.

\section*{\hspace*{-5mm}ACKNOWLEDGMENTS}
\noindent
This work has been supported by the European Research Council
(ERC) under the European Union's Horizon 2020 research and innovation programme (Grant Agreement No. 694248).


\begin{thebibliography}{99}

 \bibitem{Weyl1929}
 H. Weyl, 
 Elektron und gravitation,
 I. Z. Phys. {\bf 56}, 330--352 (1929).
 
 \bibitem{NielsenNinomiya1981}
H.B. Nielsen, M. Ninomiya:
Absence of neutrinos on a lattice.  I - Proof by homotopy theory,
Nucl. Phys. B \textbf{185}, 20  (1981);
Absence of neutrinos on a lattice. II - Intuitive homotopy proof,
Nucl. Phys. B \textbf{193}, 173 (1981).

\bibitem{FrogNielBook}
C.D. Froggatt  and  H.B. Nielsen,
 {\it Origin of Symmetry}, World Scientific, Singapore (1991).

\bibitem{GrinevichVolovik1988}
 P.G. Grinevich, G.E. Volovik,
 Topology of gap nodes in superfluid  3He:  $\pi_4$ homotopy group for 3He-B disclination,
J. Low Temp. Phys. {\bf 72}, 371  (1988).

\bibitem{Horava2005}
P. Ho\v{r}ava, Stability of Fermi surfaces and $K$-theory, Phys. Rev. Lett.
\textbf{95}, 016405 (2005).

\bibitem{Herring1937}
C. Herring,  
Accidental degeneracy in the energy bands of crystals,
 Phys. Rev. {\bf 52},  365--373  (1937).

\bibitem{Abrikosov1971}
A.A. Abrikosov and S.D. Beneslavskii,
Possible existence of substances intermediate between metals and dielectrics,
JETP {\bf 32}, 699--798 (1971).

\bibitem{Abrikosov1972}
A.A. Abrikosov,
Some properties of gapless semiconductors of the second kind,
J. Low Temp. Phys. {\bf 5}, 141--154 (1972).

\bibitem{WanEtAl2011}
X. Wan, A. M. Turner, A. Vishwanath, and S. Y. Savrasov, Topological semimetal and Fermi-arc surface states in the electronic structure of pyrochlore iridates, Phys. Rev. B {\bf 83}, 205101 (2011).

\bibitem{Burkov2011a}
A.A. Burkov and L. Balents, 
Weyl semimetal in a topological insulator multilayer,
Phys. Rev. Lett. {\bf 107}, 127205 (2011).

\bibitem{Weng2015}
Hongming Weng, Chen Fang, Zhong Fang, B. Andrei Bernevig, Xi Dai, 
Weyl semimetal phase in noncentrosymmetric transition-metal monophosphides, 
Phys. Rev. X {\bf 5}, 011029 (2015).

\bibitem{Huang2015}
Shin-Ming Huang, Su-Yang Xu, Ilya Belopolski, Chi-Cheng Lee, Guoqing Chang, BaoKaiWang,
Nasser Alidoust, Guang Bian, Madhab Neupane, Chenglong Zhang, Shuang Jia, Arun Bansil,
Hsin Lin and M. Zahid Hasan,
 A Weyl fermion semimetal with surface Fermi arcs in the transition metal monopnictide TaAs class,
 Nat. Commun. {\bf 6}, 7373 (2015).

\bibitem{Lv2015}
 B.Q. Lv, H.M. Weng, B.B. Fu, X.P. Wang, H. Miao,
J. Ma, P. Richard, X.C. Huang, L.X. Zhao, G.F. Chen,
Z. Fang, X. Dai, T. Qian, and H. Ding, 
Experimental discovery of Weyl semimetal TaAs,
Phys. Rev. X {\bf 5}, 031013 (2015).

\bibitem{Xu2015}
Su-Yang Xu, I. Belopolski, N. Alidoust, M. Neupane,
Guang Bian, Chenglong Zhang, R. Sankar, Guoqing Chang, Zhujun Yuan,
Chi-Cheng Lee, Shin-Ming Huang, Hao Zheng, Jie Ma, D.S. Sanchez,
BaoKai Wang, A. Bansil, Fangcheng Chou, P.P. Shibayev, Hsin Lin,
Shuang Jia, M. Zahid Hasan,
Discovery of a Weyl fermion semimetal and topological Fermi arcs,
Science {\bf 349}, 613--617 (2015).

\bibitem{Lu2015}
Ling Lu, Zhiyu Wang, Dexin Ye, Lixin Ran, Liang Fu, John D. Joannopoulos, Marin Soljacic,
Experimental observation of Weyl points,
Science {\bf 349}, 622--624 (2015).

\bibitem{ChangEtAl2016}
Guoqing Chang, Su-Yang Xu, Daniel S. Sanchez, Shin-Ming Huang, Chi-Cheng Lee, Tay-Rong Chang, Guang Bian, Hao Zheng, Ilya Belopolski, Nasser Alidoust, Horng-Tay Jeng,
Arun Bansil, Hsin Lin, M. Zahid Hasan, 
A strongly robust type II Weyl fermion semimetal state in Ta$_3$S$_2$, 
Sci. Adv. 2016; 2 : e1600295 (2016).

\bibitem{SunEtAl2015}
Yan Sun, Shu-Chun Wu, Mazhar N. Ali, Claudia Felser, and Binghai Yan, Prediction of Weyl semimetal in orthorhombic MoTe$_2$, Phys. Rev. B {\bf 92}, 161107(R) (2015).

\bibitem{XuEtAl2016b}
N. Xu, H. M. Weng, B. Q. Lv, C. Matt, J. Park, F. Bisti, V. N. Strocov, D. Gawryluk, E. Pomjakushina, K. Conder, N. C. Plumb, M. Radovic, G. Aut\`es, O. V. Yazyev, Z. Fang, X. Dai, G. Aeppli, T. Qian, J. Mesot, H. Ding, M. Shi, Observation of Weyl nodes and Fermi arcs in tantalum phosphide, Nature Communications {\bf 7}, 11006 (2016).

\bibitem{KoepernikEtAl2016}
K. Koepernik, D. Kasinathan, D. V. Efremov, Seunghyun Khim, Sergey Borisenko, Bernd B\"uchner, and Jeroen van den Brink, TaIrTe$_4$: A ternary type-II Weyl semimetal,
Phys. Rev. B 93, 201101(R) (2016); Seunghyun Khim, Klaus Koepernik, Dmitry V. Efremov, J. Klotz, T. F\"orster, J. Wosnitza, Mihai I. Sturza, Sabine Wurmehl, Christian Hess, Jeroen van den Brink, and Bernd B\"uchner, Magnetotransport and de Haas--van Alphen measurements in the type-II Weyl semimetal 
TaIrTe$_4$, Phys. Rev. B {\bf 94}, 165145 (2016).

\bibitem{XuEtAl2016}
Su-Yang Xu, Nasser Alidoust, Guoqing Chang, Hong Lu, Bahadur Singh, Ilya Belopolski, Daniel Sanchez, Xiao Zhang, Guang Bian, Hao Zheng, Marius-Adrian Husanu, Yi Bian, Shin-Ming Huang, Chuang-Han Hsu, Tay-Rong Chang, Horng-Tay Jeng, Arun Bansil, Vladimir N. Strocov, Hsin Lin, Shuang Jia, M. Zahid Hasan,
Discovery of Lorentz-violating Weyl fermion semimetal state in LaAlGe materials, 
arXiv:1603.07318 [cond-mat.mes-hall] (2016).

\bibitem{WangEtAl2016}
Zhijun Wang, Dominik Gresch, Alexey A. Soluyanov, Weiwei Xie, S. Kushwaha, Xi Dai, Matthias Troyer, Robert J. Cava, and B. Andrei Bernevig, MoTe$_2$: A Type-II Weyl Topological Metal, Phys. Rev. Lett. {\bf 117}, 056805 (2016).

\bibitem{TamaiEtAl2016}
A. Tamai, Q.S. Wu, I. Cucchi, F.Y. Bruno, S. Ricc\`o, T.K. Kim, M. Hoesch, C. Barreteau, E. Giannini, C. Besnard, A.A. Soluyanov, and F. Baumberger, Fermi Arcs and Their Topological Character in the Candidate Type-II Weyl Semimetal MoTe$_2$, Phys. Rev. X 6, 031021 (2016).

\bibitem{JiangEtAl2016}
J. Jiang, Z. K. Liu, Y. Sun, H. F. Yang, R. Rajamathi, Y. P. Qi, L. X. Yang, C. Chen, H. Peng, C.-C. Hwang, S. Z. Sun, S.-K. Mo, I. Vobornik, J. Fujii, S. S. P. Parkin, C. Felser, B. H. Yan, Y. L. Chen, Observation of the Type-II Weyl Semimetal Phase in MoTe2, Nature Communications {\bf 8}, 13973 (2017).

\bibitem{HuangZhouDuan2016}
Huaqing Huang, Shuyun Zhou, and Wenhui Duan, Type-II Dirac fermions in the PtSe$_2$ class of transition metal dichalcogenides, Phys. Rev. B {\bf 94}, 121117(R) (2016).

\bibitem{LeEtAl2016}
Congcong Le, Shengshan Qin, Xianxin Wu, Xia Dai, Peiyuan Fu, Jiangping Hu, Three-dimensional Critical Dirac semimetal in KMgBi, arXiv:1606.05042 (2016).

\bibitem{ChangEtAl2016b}
Tay-Rong Chang, Su-Yang Xu, Daniel S. Sanchez, Shin-Ming Huang, Guoqing Chang, Chuang-Han Hsu, Guang Bian, Ilya Belopolski, Zhi-Ming Yu, Xicheng Xu, Cheng Xiang, Shengyuan A. Yang, Titus Neupert, Horng-Tay Jeng, Hsin Lin, M. Zahid Hasan, Type-II Topological Dirac Semimetals: Theory and Materials Prediction (VAl3 family), arXiv:1606.07555 (2016).

\bibitem{Bevan1997}
T.D.C. Bevan, A.J. Manninen, J.B. Cook, J.R. Hook, H.E. Hall, T. Vachaspati and G.E. Volovik,
Momentum creation by vortices in superfluid $^3$He as a model of primordial baryogenesis,
Nature  {\bf 386},  689-692 (1997).

\bibitem{Volovik2003}
G.E. Volovik,
The Universe in a Helium Droplet,
Clarendon Press,  Oxford (2003)

\bibitem{NeumannWigner1929} 
J. von Neumann and E. Wigner,
\"Uber das Verhalten von Eigenwerten bei adiabatischen Prozessen,
Phys. Z. {\bf 30}, 467 (1929).

\bibitem{Novikov1981}
S.P. Novikov, 
 Magnetic Bloch functions and vector bundles. Typical dispersion laws and their quantum numbers, 
 Sov. Math., Dokl. {\bf 23}, 298--303 (1981).

\bibitem{VolovikZubkov2014}
G.E. Volovik and M.A. Zubkov, 
Emergent Weyl spinors in multi-fermion systems, 
Nuclear Physics B {\bf 881}, 514  (2014).

%\bibitem{XuZhangZhang15}
\bibitem{YongXu2015}
Y. Xu, F. Zhang, and C. Zhang,
Structured Weyl points in spin-orbit coupled fermionic superfluids,
Phys. Rev. Lett. \textbf{115}, 265304 (2015).

\bibitem{Soluyanov2015}
A.A. Soluyanov, D. Gresch, Zhijun Wang, QuanSheng Wu, M. Troyer, Xi Dai, B.A. Bernevig,  
Type-II Weyl semimetals,
Nature {\bf 527}, 495--498 (2015).

\bibitem{HuhtalaVolovik2002}
P. Huhtala and  G.E. Volovik,  
Fermionic microstates within Painlev\'e-Gullstrand black hole, 
ZhETF {\bf 121}, 995-1003; JETP {\bf 94}, 853-861 (2002); gr-qc/0111055.

\bibitem{Volovik2016}
G.E. Volovik,
Black hole and Hawking radiation by type-II Weyl fermions,
Pis'ma ZhETF {\bf 104},  660--661 (2016),
JETP Lett.  {\bf 104},  645--648 (2016),
arXiv:1610.00521; K. Zhang, G.E. Volovik, Lifshitz transitions via the type-II Dirac and type-II Weyl points, arXiv:1604.00849 (2016).

\bibitem{ADM}
R. Arnowitt, S. Deser, C. W. Misner,
The Dynamics of General Relativity,
p. 227 in \emph{Gravitation: an introduction to current
research}. Edited by Louis Witten. John Wiley \& Sons Inc., New York, London, (1962); Gen. Relativ. Gravit. 40:1997--2027  (2008).

\bibitem{Frolov2004}
V.P. Frolov and K.A. Stevens
Stationary strings near a higher-dimensional rotating black hole,
Phys. Rev. D {\bf 70}, 044035 (2004).

\bibitem{Diakonov2011}
D. Diakonov,
Towards lattice-regularized Quantum Gravity,
arXiv:1109.0091 (2011).

\bibitem{Diakonov2012}
A.A. Vladimirov, D. Diakonov,
Phase transitions in spinor quantum gravity on a lattice,
Phys. Rev. D {\bf 86}, 104019 (2012). 

\bibitem{Bonnor2005}
W.B. Bonnor and  B.R. Steadman,
Exact solutions of the Einstein-Maxwell equations with closed timelike curves,
Gen. Rel. Grav.
{\bf  37},1833--1844 (2005).

\bibitem{Kajari2004}
E. Kajari, R. Walser, W. P. Schleich, A. Delgado, 
Sagnac Effect of G\"odel's Universe,
Gen. Rel. Grav. {\bf 36}, 2289 (2004).

\bibitem{Volovik2009}
G.E. Volovik, 
Topological invariant  for superfluid  $^3$He-B and quantum phase transitions,
Pis'ma ZhETF {\bf 90}, 639--643 (2009);  JETP Lett. {\bf 90}, 587--591 (2009);
arXiv:0909.3084.

\bibitem{Volovik2010}
 G.E. Volovik, 
Topological invariants  for Standard Model: from semi-metal to topological insulator,
 Pis'ma ZhETF {\bf 91}, 61--67 (2010);   JETP Lett. {\bf 91}, 55--61 (2010);
arXiv:0912.0502.

\bibitem{VayrynenVolovik2011}
J.I. V\"ayrynen and G.E. Volovik,
Soft topological objects in topological media, 
Pis'ma ZhETF {\bf 93}, 378--382 (2011); JETP Lett. {\bf 93}, 344-348 (2011);
arXiv:1101.1179.

\bibitem{Wang2013}
Z. Wang and B. Yan, 
Topological Hamiltonian as an exact tool for topological invariants,
J. Phys. Condens. Matter {\bf 25}, 155601 (2013).

\bibitem{Abanin2014}
W. Witczak-Krempa, M. Knap  and D. Abanin,
Interacting Weyl semimetals: Characterization via the topological Hamiltonian
and its breakdown,
Phys. Rev. Lett.  {\bf 113}, 136402 (2014).

\end{thebibliography}
\end{document}